\documentclass[12pt]{iopart}
\usepackage{graphicx}
\usepackage{bm}
\usepackage{color}

\begin{document}

\title[]{Full-zone Persistent Spin Textures with Giant Spin Splitting in Two-dimensional  Group IV-V Compounds}

\author{Moh. Adhib Ulil Absor$^{1,*}$, Arif Lukmantoro$^{1}$, and Iman Santoso$^{1}$}
\address{$^{1}$Department of Physics, Universitas Gadjah Mada, Sekip Utara BLS 21 Yogyakarta 55281, Indonesia.}
\ead{$^{*}$adib@ugm.ac.id}

\vspace{10pt}
\begin{indented}
\item[]February 2021
\end{indented}

\begin{abstract}
Persistent spin texture (PST), a property of solid-state materials maintaining unidirectional spin polarization in the momentum $k$-space, offers a route to deliver the necessary long carrier spin lifetimes through the persistent spin helix (PSH) mechanism. However, most of the discovered PST locally occurred in the small part around certain high symmetry $k$-points or lines in the first Brillouin zone (FBZ), thus limiting the stability of the PSH state. Herein, by symmetry analysis and first-principles calculations, we report the emergence of full-zone PST (FZPST), a phenomenon displaying the PST in the whole FBZ, in the two-dimensional group IV-V $A_{2}B_{2}$ ($A$ = Si, Sn, Ge; $B$ = Bi, Sb) compounds. Due to the existence of the in-plane mirror symmetry operation in the wave vector point group symmetry for the arbitrary $\vec{k}$ in the whole FBZ, fully out-of-plane spin polarization is observed in the $k$-space, thus maintaining the FZPST. Importantly, we observed giant spin splitting in which the PST sustains, supporting large SOC parameters and small wavelengths of the PSH states. Our $\vec{k}\cdot\vec{p}$ analysis demonstrated that the FZPST is robust for the non-degenerate bands, which can be effectively controlled by the application of an external electric field, thus offering a promising platform for future spintronic applications.
\end{abstract}

%
%
%
%
%

\section{Introduction}

The study of relativistic effects in solids, i.e., the spin-orbit coupling (SOC), has attracted increasing interest in the field of spintronics since it allows for manipulation of an electron's spin degree of freedom without additional external magnetic field \cite{Manchon}. Many intriguing SOC-related phenomena were observed, including spin relaxation \cite{Fabian, Averkiev}, spin Hall effect \cite{Qi}, spin galvanic effect \cite{Ganichev}, and spin ballistic transport \cite{Lu}. In a system lacking inversion symmetry, the SOC induces momentum ($k$)-dependent spin-orbit field (SOF), $\vec{\Omega}(\vec{k})\propto \vec{E}\times\vec{p}$, where $\vec{E}$ is the electric field originated from the crystal inversion asymmetry and $\vec{p}$ is the momentum, that lifts Kramer’s spin degeneracy and leads to the nontrivial $k$-dependent spin textures of the spin-split bands through the so-called Rashba \cite{Rashba} and Dresselhaus \cite{Dress} effects. In particular, the Rashba effect attracted much attention since it can be manipulated electrically to produce non-equilibrium spin polarization \cite{Nitta, Kuhlen}, which plays an important role in the spin-field effect transistors (SFET) \cite{Datta}. 

Practically, materials with a strong Rashba effect are desirable in spintronic applications since a large Rashba SOC parameter is in favor of room temperature device operations \cite{Xu2014}. However, the strong Rashba SOC is also known to induce the undesired effect of causing spin decoherence, which plays a detrimental role in the spin lifetime \cite{Dyakonov}. Due to the $k$-dependent SOF, electron scatterings by defect and impurity randomize the spin in a diffusive transport regime, which induces the fast spin decoherence through the Dyakonov-Perel spin relaxation mechanism \cite{Dyakonov}, and hence limits the performance of the spintronics functionality. This problem can further be resolved by designing the materials supporting unidirectional SOF. Under the unidirectional SOF, the spin polarization is unidirectionally oriented in the $k$-space, resulting in the so-called persistent spin texture (PST) \cite{Schliemann, SchliemannJ}, thus enabling long-range spin transport without dissipation through persistent spin helix (PSH) mechanism \cite{Bernevig, SchliemannJ, Altmann}. Previously, GaAs/AlGaAs \cite{Koralek2009, Walser2012} and InGaAs/InAlAs \cite{Kohda, Kohda_a, Sasaki2014} quantum well (QW) heterostructures have been reported to demonstrate the emergence of the PSH state arising from a balance between the strength of the Rashba and Dresselhaus SOCs; however, these artificial structures require atomic precision by tuning the QW width and carefully controlled carrier densities through doping level and applied external electric field.   

Recently, a different approach for achieving the PST has been proposed, imposing the specific symmetry of the crystal for the systems having a crystallographic polar axis to preserve the unidirectional SOF \cite{Tao2018}. While it remains to be confirmed experimentally, this approach allows us to design the PST without requiring the fine-tuning between the Rashba and Dresselhaus SOCs. This is particularly achieved on the class of materials exhibiting intrinsic spontaneous electric polarization as previously reported on bulk ferroelectric BiInO$_{3}$ \cite{Tao2018}, CsBiNb$_{2}$O$_{7}$ \cite{Autieri2019}, layered ferroelectric oxide Bi$_{2}$WO$_{6}$ \cite{Djani}, and several two-dimensional (2D) ferroelectric systems including WO$_{2}$Cl$_{2}$ \cite{Ai2019}, Ga$XY$($X$=Se, Te; $Y$=Cl, Br, I) \cite{Absor2021a, Absor2021b}, hybrid perovskite benzyl ammonium lead-halide \cite {Jia}, and group-IV monochalcogenide \cite{Absor2021c, Absor2019a, Anshory2020, Absor2019b, Lee2020}. More recently, the symmetry-protected PST with purely cubic spin splitting has also been reported in bulk materials crystallizing in the $\bar{6}m2$ and $\bar{6}$ point groups \cite{Zhao2020}. In addition, the PST driven by the lower symmetry of the crystal has also been reported on wurtzite ZnO [10$\bar{1}$0] surface \cite{Absor2015} and several 2D transition metal dichalcogenides (TMDCs) with the line defect \cite{Li2019, Absor2020}. Although the symmetry-enforced PSTs have been extensively studied, most of them locally occurred in the small part around the high symmetry $\vec{k}$ points or lines in the first Brillouin zone (FBZ) \cite{Tao2018, Autieri2019, Ai2019, Absor2021a, Absor2021b, Absor2021c, Absor2019a, Anshory2020, Absor2019b, Lee2020, Zhao2020}. This implies that the spin deviation away from the uniaxial SOF appears for the larger region in the FBZ, which has a considerable effect to induce the significant spin scattering and hence limits the stability of the PSH states \cite{Tao2018, LU2020}. Therefore, finding novel structures supporting the PST covering a substantially large region in the FBZ is highly desirable, which is important to produce a stable PSH state without controlling the Fermi level in the specific FBZ region.

In this paper, through symmetry analysis and first-principles density-functional theory (DFT) calculations, we show that the PST could exist in the whole FBZ of the 2D group IV-V $A_{2}B_{2}$ ($A$ = Si, Sn, Ge; $B$ = Bi, Sb) compounds. We dub this phenomenon the full-zone PST (FZPST). We found that the FZPST observed in the present systems is characterized by fully out-of-plane spin polarization in the $k$-space, which is enforced by the in-plane mirror symmetry operation in the wave vector point group symmetry for the arbitrary $\vec{k}$ in the whole FBZ. More importantly, the observed FZPST exhibits giant spin splitting, which is particularly visible around the $M$ points in the conduction band minimum, thus resulting in large SOC parameters and small wavelengths of the PSH states. Our $\vec{k}\cdot\vec{p}$ analysis derived from the invariant method \cite{Winkler, Vajna} confirmed that this FZPST is robust for the non-degenerate bands, which can be sensitively tuned by an external electric field. Finally, a possible application of the FZPST in the present system for spintronic devices will be highlighted.

\section{Computational Details}

Our DFT calculations were carried out based on norm-conserving pseudo-potentials and optimized pseudo-atomic localized basis functions \cite{Troullier} implemented in the OPENMX code \cite{OpenmX, Ozaki, Ozakikinoa, Ozakikinoa}. The exchange-correlation functional was treated within generalized gradient approximation by Perdew, Burke, and Ernzerhof (GGA-PBE) \cite{gga_pbe, Kohn}. The basis functions were expanded by the linear combination of multiple pseudo atomic orbitals generated using a confinement scheme \cite{Ozaki, Ozakikino, Ozakikinoa}, where two $s$-, two $p$-, two $d$-character numerical pseudo atomic orbitals were used. We applied a periodic slab to model the $A_{2}B_{2}$ monolayers (MLs), where a sufficiently large vacuum layer (35 \AA) was used to avoid the spurious interaction between the slabs [Fig. 1(a)]. The $12\times12\times1$ $k$-point mesh was used to discretize the FBZ [Fig. 1(b)]. The energy convergence criterion was set to $10^{-9}$ eV during the structural relaxation. The lattice and positions of the atoms were optimized until the Hellmann-Feynman force components acting on each atom were less than 1 meV/\AA. The vibrational properties and the phonon dispersion relations were acquired via the small-displacement method using the PHONOPY code \cite{TOGO2015}. Ab-initio molecular dynamics (AIMD) simulations were also carried out to examine the thermal stability of the $A_{2}B_{2}$ MLs by using a 4×4×1 supercell consisting of 64 atoms at room temperature (300 K) with a total simulation time of 6 ps. 

The spin textures in the $k$-space were evaluated by using the spin density matrix of the spinor wave functions obtained from the DFT calculations \cite{Kotaka}. As we applied previously on various 2D materials \cite{Absor2021a, Absor2021b, Absor2021c, Absor2020, Absor2019a, Absor2019b, Absor2018, Absor2017}, we deduced the spin vector component ($S_{x}$, $S_{y}$, $S_{z}$) of the spin polarization in the reciprocal lattice vector $\vec{k}$ from the spin density matrix, $P_{\sigma \sigma^{'}}(\vec{k},\mu)$, calculated using the following relation,  
\begin{equation}
\label{1}
P_{\sigma \sigma^{'}}(\vec{k},\mu)=\int \Psi^{\sigma}_{\mu}(\vec{r},\vec{k})\Psi^{\sigma^{'}}_{\mu}(\vec{r},\vec{k}) d\vec{r}         \end{equation}
where $\sigma$ ($\sigma^{'}$) is the spin index ($\uparrow$ or $\downarrow$) and $\mu$ is the band index. Here, $\Psi^{\sigma}_{\mu}(\vec{r},\vec{k})$ is the spinor Bloch wave function, which is obtained from the OpenMX calculations after self-consistent is achieved. 

\section{Results and Discussion}

\subsection{Symmetry-enforced FZPST in the 2D non-magnetic systems}

First, we start our discussion by analyzing the FZPST from the symmetry perspective. It has been generally understood that the classification of the materials having spin-orbit-induced spin splitting and spin polarization is usually determined by the crystallography point group symmetry (CPGS) such as polar and non-polar crystal classes. However, for a fixed CPGS, a journey through the FBZ from one $\vec{k}$ point to another leads to a change in the wave vector point group symmetry (WPGS). Consequently, a significant change of the spin textures, defined as the expectation value of the spin polarization vector $\vec{S}(\vec{k})$ in the $k$-space, may occur \cite{Acosta}. Since the $\vec{S}(\vec{k})$ is bound to be parallel to the rotational axis and perpendicular to the mirror plane, the symmetry operation types in the WPGS impose symmetry restriction on $\vec{S}(\vec{k})$. Therefore, the FZPST is achieved as long as the WPGS keeps the unidirectional spin polarization $\vec{S}(\vec{k})$ invariant in the whole FBZ. To realize the symmetry-protected FZPST, there must exist a symmetry operator of the WPGS protecting the wave vector $\vec{k}$ invariant in the whole FBZ. By defining $\hat{\Omega}$ as the symmetry operator of the WPGS, the representation of $\hat{\Omega}$, $D(\hat{\Omega})$, can be determined through the following relation, $\hat{\Omega}\vec{S}(\vec{k})=\vec{S}(\vec{k})D(\hat{\Omega})$. Therefore, the allowed spin polarization direction, $\vec{S}(\vec{k})=(S_{x}, S_{y}, S_{z})$, at given $\vec{k}$ can be obtained by the invariant condition of the $\vec{S}(\vec{k})$ under the symmetry opertor $\hat{\Omega}$ as follows, $\vec{S}(\vec{k}) D(\hat{\Omega})=\vec{S}(\vec{k})$.

Now, let us consider the 2D crystalline systems where the out-of-plane direction is chosen to be oriented along the $z$-axis. In this case, the FBZ is a 2D plane with $k_{z}=0$. For this 2D system, we find that only the identity $E$ and in-plane mirror symmetry $M_{xy}$ operations maintain the $\vec{k}$ point being invariant in the whole FBZ. Accordingly, the allowed $\vec{S}(\vec{k})$ under the restriction of these symmetry operations enforced to unidirectionally oriented along the out-of-plane ($z$) direction in the whole FBZ through the following transformation, $(S_{x}, S_{y}, S_{z})\stackrel{M_{xy}}{\rightarrow}(-S_{x}, -S_{y}, S_{z})$, leading to the FZPST. We emphasized here that for non-magnetic 2D crystalline systems, all the $\vec{k}$ points are invariant under time-reversal symmetry operation $\mathcal{T}$. For a given inversion symmetry operation $\mathcal{P}$, we can construct an anti-unitary operator, $\mathcal{P}\mathcal{T}$, which flips the spin polarizatrion vector $\vec{S}=(S_{x}, S_{y}, S_{y})$, so that $\left\langle \vec{S}_{\vec{k}}\right\rangle=0 $ for any $\vec{k}$ in the FBZ. Therefore, we conclude that the FZPST could exist in the 2D non-magnetic systems if and only if the wave vector $\vec{k}$ in the whole FBZ has the in-plane mirror symmetry $M_{xy}$ but without inversion symmetry $\mathcal{P}$.  

\begin{table*}
\caption{Summarization of all the CPGS in the 2D systems containing (Y) or not containing (N) in-plane mirror symmetry $M_{xy}$ and inversion symmetry $\mathcal{P}$. The last line refers to whether the FZPST exist (Y) or not (N) in the 2D FBZ, which is determined by considering the WPGS of $\vec{k}$ in the whole 2D FBZ. } 
\begin{tabular}{c c c c c c c c c c c c c c c} 
\hline\hline 
            & $C_{1}$ & $C_{i}$  & $C_{2}$ & $C_{s}$ & $C_{2h}$  & $C_{2v}$   & $C_{3}$ & $C_{3i}$  & $C_{3h}$ & $C_{3v}$  & $C_{4}$               & $C_{4h}$ & $C_{4v}$  & $C_{6}$ \\ 
\hline 
 $M_{xy}$   & N & N & N & Y & Y  &  Y  & N & N  & Y & N  & N  & Y  & N  & N  \\
	$\mathcal{P}$	      & N & Y & N & N & Y  &  N  & N & N  & N & N  & N  & Y  & N  & N  \\
	FZPST    	& N & N & N & Y & N  &  Y  & N & N  & Y & N  & N  & N  & N  & N  \\
		        &   &   &   &   &    &     &   &   &  &   &   &   &   &   \\
		        & $C_{6h}$  &   $C_{6v}$ & $D_{2}$ & $D_{2h}$ & $D_{2d}$ & $D_{3}$ & $D_{3h}$  & $D_{3d}$   & $D_{4}$  & $D_{4h}$  & $D_{6}$ & $D_{6h}$  & $S_{4}$  &  \\
\hline 
	$M_{xy}$	& Y & N & N & Y & N & N & Y  &  N  & N & Y  & N & Y  &  N &   \\
	$\mathcal{P}$	  & Y & N    & N & Y & N & N & N  &  Y  & N & Y  & N & Y  &  N &    \\
	FZPST	   & N & N & N & N & N & N & Y  &  N  & N & N  & N & N  &  N &    \\
\hline\hline 
\end{tabular}
\label{table:Table 1} 
\end{table*}

Finally, we search the possible FZPST in the 2D crystalline system identified using the criteria whether the CPGS containing (Y) or not containing (N) both the in-plane mirror symmetry $M_{xy}$ and inversion symmetry ($\mathcal{P} $) operations under the guidance of Table 1. When the CPGS contains the in-plane mirror symmetry $M_{xy}$ operation, the WPGS for the $\vec{k}$ point in the whole FBZ should be invariant under this symmetry operation. Therefore, we find that the following CPGS of the 2D crystalline systems is possible to maintain the FZPST including $C_{s}$, $C_{2v}$, $C_{3h}$, and $D_{3h}$. 

In the next section, we discuss our results from the first-principles DFT calculations on the group IV-V $A_{2}B_{2}$ ML compounds to confirm the above-predicted FZPST.    

\subsection{FZPST in 2D group IV-V $A_{2}B_{2}$ compounds}

\subsubsection{Structural symmetry, stability, and electronic properties.}

It has been previously reported that a combination between group IV elements, $A$ ($A$ = Si, Sn, Ge), and group V elements, $B$ ($B$ = Bi, Sb), forms a stable 2D layered $A_{2}B_{2}$ compounds \cite{Ozdamar2018, Bafekry2021, Bafekry2020}. These compounds consists of covalently bonded quadruple atomic layers in a $B-A-A-B$ sequence, forming a trigonal prismatic structure because $A$ atoms are coordinated in the form of a triangular prism with respect to the $B$ dimer, similar to the 2H-phase of the TMDCs MLs \cite{Zhu2011, Absor2016, Absor2017, Sun2021}. The atomic structure and corresponding to the FBZ are schematically shown in Figs. 1(a) and (b), respectively. From symmetry point of view, the crystal symmetry of the $A_{2}B_{2}$ ML compounds has $P\bar{6}m_{2}$ space group. There are six classes of the symmetry operation making the structure of the $A_{2}B_{2}$ MLs invariant including identity ($E$), in-plane mirror reflection ($M_{xy}$), out-of-plane mirror reflection ($M_{xz}$, $M_{xz}^{'}$, $M_{xz}^{"}$), two-fold rotation around the axis parallel to the $M_{xz}$ plane ($C_{2}$, $C_{2}^{'}$, $C_{2}^{"}$), three-fold rotation around the $z$-axis ($C_{3}$, $C^{2}_{3}$), and three-fold improper rotation around the $z$-axis ($S_{3}$, $S^{2}_{3}$) [see Fig. 1(a)]. Therefore, the CPGS of the $A_{2}B_{2}$ MLs belongs to $D_{3h}$ point group. The optimized structural-related parameters including lattice constant $a$, bond distances, and bond angles in the $A_{2}B_{2}$ MLs are presented in Table SI in the the Supporting Information \cite{Supporting}, and overall are in a good agreement with previous results \cite{Bafekry2021, Bafekry2020, Ozdamar2018}.

\begin{figure}
	\centering		
	\includegraphics[width=1.0\textwidth]{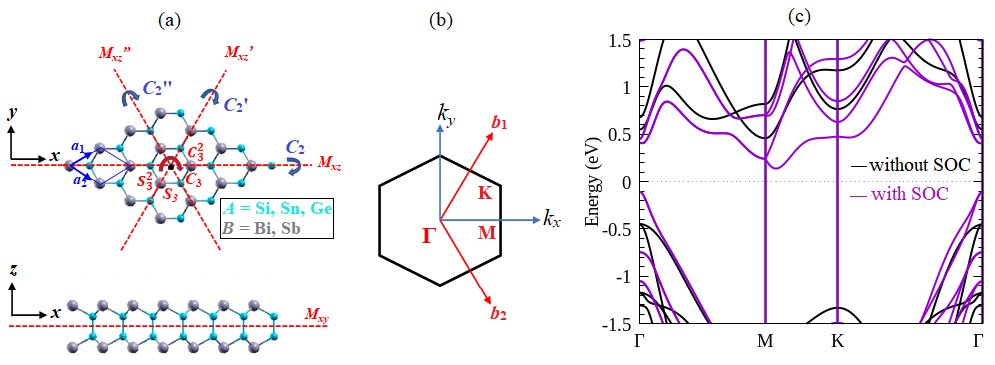}
	\caption{(a) Atomic structures of the $A_{2}B_{2}$ MLs ($A$ = Si, Ge, Sn; $B$ = Sb, Bi) corresponding to the first Brillouin zone (FBZ) (b) are shown. The unit cell of the crystal is indicated by the blue lines. The FBZ is characterized by the $\Gamma$, $X$, $Y$, and $M$ high symmetry points. The symmetry operations in the crystal consisting of identity ($E$), in-plane mirror reflection ($M_{xy}$), out-of-plane mirror reflection ($M_{xz}$, $M_{xz}^{'}$, $M_{xz}^{"}$), two-fold rotation around the axis parallel to the $M_{xz}$ plane ($C_{2}$, $C_{2}^{'}$, $C_{2}^{"}$), three-fold rotation around the $z$-axis ($C_{3}$, $C^{2}_{3}$), and three-fold improper rotation around the $z$-axis ($S_{3}$, $S^{2}_{3}$), are shown. (c) Electronic band structure of the Si$_{2}$Bi$_{2}$ calculated without (black lines) and with (magenta line) the SOC is shown.}
	\label{figure:Figure1}
\end{figure}

To confirm the stability of the $A_{2}B_{2}$ MLs, several assessments are carried out including cohesive energy ($E_{coh}$), phonon dispersion bands, and the AIMD simulation. All the calculated results confirmed that the optimized $A_{2}B_{2}$ ML compounds are energetically, dynamically, and thermally stable; see the Supporting Information \cite{Supporting}. It has been reported previously that several layered $A_{2}B_{2}$ compounds ($A$ = Si, Ge; $B$ = As, P) have been experimentally synthesized in the bulk form \cite{Barreteau}, which opens an opportunity for the realization of these compounds in the 2D form. Considering the similar structural symmetry and stability, the experimental realization of the $A_{2}B_{2}$ MLs ($A$= Si, Ge, Sn; $B$= Sb, Bi) is highly plausible.  

\begin{figure}
	\centering		
	\includegraphics[width=0.8\textwidth]{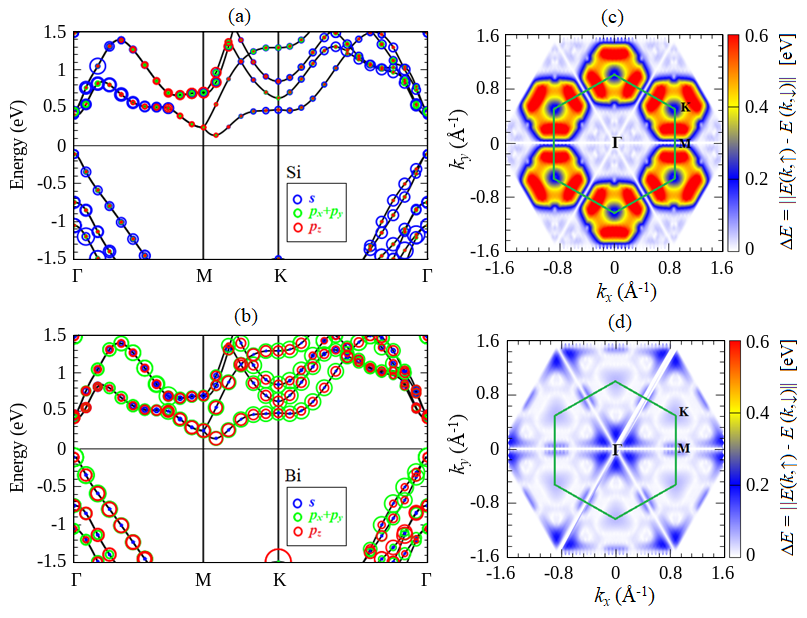}
	\caption{Orbital-resolved electronic band structures of the Si$_{2}$Bi$_{2}$ ML calculated for: (a) Si and (b) Bi atoms are presented. Here, blue, green, and red colors of circle represent $s$, $(p_{x}+p_{y})$, and $p_{z}$ orbitals. The radii of the circles reflect the magnitudes of spectral weight of the particular orbitals to the band.(c)-(d) Spin-splitting energy of the Si$_{2}$Bi$_{2}$ ML mapped in the first Brillouin zone for CBM and VBM, respectively, are shown. The magnitude of the spin-splitting energy, $\Delta E$, defined as $\Delta E=|E(k,\uparrow)-E(k,\downarrow)|$, where $E(k,\uparrow)$ and $E(k,\downarrow)$ are the energy bands with up spin and down spin, respectively, is represented by the color scales.}
	\label{figure:Figure2}
\end{figure}

To characterize the electronic properties of the $A_{2}B_{2}$ MLs, we show in Fig. 1(c) the calculated band structure of the Si$_{2}$Bi$_{2}$ ML as a representative example, while the band structure for other members of the $A_{2}B_{2}$ MLs are presented in Figure S3 in the Supporting Information \cite{Supporting}. Consistent with previous reports \cite{Ozdamar2018, Bafekry2021, Bafekry2020}, the Si$_{2}$Bi$_{2}$ ML is an indirect semiconductor where the valence band maximum (VBM) is located on the $\Gamma$ point, while the conduction band minimum (CBM) lies on the $M$ point. When the SOC is taken into account, the CBM resides between $M$ and $K$ points, thus maintaining the indirect bandgap. Our calculated orbital-resolved the band projected to the atoms confirmed that the CBM is mainly originated from the contribution of  Bi-$p_{x}+p_{y}$ and Bi-$p_{z}$ orbitals, while the strong admixture of Bi-$p_{x}+p_{y}$, Bi-$p_{z}$, and Si-$s$ orbitals contributes significantly to the VBM [Figs. 2(a)-(b)].

Introducing the SOC induces a sizable band splitting due to the inversion symmetry breaking of the crystal [Fig. 1(c)]. This splitting is particularly visible for $\vec{k}$ points and lines located in the proximity of the band edges near the Fermi level, except for time-reversal-invariant $k$ points at $\Gamma$ and $M$ points, and high symmetry line at the $\Gamma-M$ line. To quantify the spin-split bands, we show in Figs. 2(c)-(d) the spin-splitting energy of the CBM and VBM mapped along the entire region of the FBZ. Consistent with the band degeneracy along the $\Gamma-M$ line, zero spin splitting is clearly visible for $\theta=n\pi/3$, where $n=\in N_{0}$ and $\theta$ is measured from $k_{y}=0$. Interestingly, we identify a strongly anisotropic spin-splitting around the $M$ point in the CBM where the maximum spin-splitting energy up to 0.72 eV is observed along the $M-K$ line [Fig. 2(c)]. This giant spin splitting is indeed larger than that observed on several TMDCs MLs [0.15 eV–0.55 eV] \cite{Zhu2011, Absor2016, Absor2017, Yao2017, Absor2017a}, which is mainly originated from the strong admixture of the in-plane (Bi-$p_{x}+p_{y}$, Si-$p_{x}+p_{y}$) and out-of-plane (Bi-$p_{z}$, Si-$p_{z}$) orbitals [Figs. 2(a)-(b)]. Remarkably, The giant spin splitting observed in the present system is certainly sufficient to ensure the proper function of spintronic devices operating at room temperature \cite{Xu2014}.

\subsubsection{DFT analysis of the FZPST.}

Now, we discuss the emergence of the FZPST in the Si$_{2}$Bi$_{2}$ ML obtained from the DFT calculation. The FZPST can be observed from the spin polarization of the spin-split bands in the whole FBZ. However, before we show our DFT results, we first clarify the existence of the FZPST by identifying the WPGS of the $\vec{k}$ in the whole FBZ. As previously mentioned that the CPGS of the $A_{2}B_{2}$ MLs belongs to the $D_{3h}$ point group, thus the WPGS of the $\vec{k}$ in the FBZ should contain some of the symmetry operations in the $D_{3h}$ point group. In Table 2, we list the possible WPGS of the $\vec{k}$ corresponding to the symmetry operations at the high symmetry points and lines in the FBZ. We find that the WPGS of the $\vec{k}$ in the whole FBZ belongs to $C_{s}$ point group containing $E$ and $M_{xy}$ symmetries, thus the FZPST should exist in the $A_{2}B_{2}$ MLs having fully out-of-plane spin polarization.

\begin{table*}
\caption{The WPGS of the $\vec{k}$ in the high symmetry points and lines in the FBZ for the $A_{2}B_{2}$ MLs. Here, the FBZ is schematically given in Fig. 1(b). } 
\begin{tabular}{ccc  ccc  ccc } 
\hline\hline 
  $\vec{k}$ point &&& WPGS  &&&  symmetry operations \\ 
\hline 
$\Gamma$ &&&  $D_{3h}$  &&&     $E$, $M_{xy}$, ($M_{xz}$, $M_{xz}^{'}$, $M_{xz}^{"}$), ($C_{2}$, $C_{2}^{'}$, $C_{2}^{"}$), ($C_{3}$, $C^{2}_{3}$), ($S_{3}$, $S^{2}_{3}$)  \\
     $M$ &&&  $C_{2v}$  &&&      $E$, $M_{xy}$, $M_{xz}$, $C_{2}$\\
     $K$ &&&  $C_{3h}$  &&&     $E$, $M_{xy}$, ($C_{3}$, $C^{2}_{3}$), ($S_{3}$, $S^{2}_{3}$) \\
		 $\Gamma-M$ &&&   $C_{2v}$  &&&  $E$, $M_{xy}$, $M_{xz}$, $C_{2}$   \\
		$\Gamma-K$ &&& $C_{s}$   &&&  $E$, $M_{xy}$   \\
		$M-K$ &&&  $C_{s}$  &&& $E$, $M_{xy}$    \\
		Whole FBZ &&&  $C_{s}$  &&&     $E$, $M_{xy}$\\
\hline\hline 
\end{tabular}
\label{table:Table 2} 
\end{table*}

\begin{figure*}
	\centering
		\includegraphics[width=1.1\textwidth]{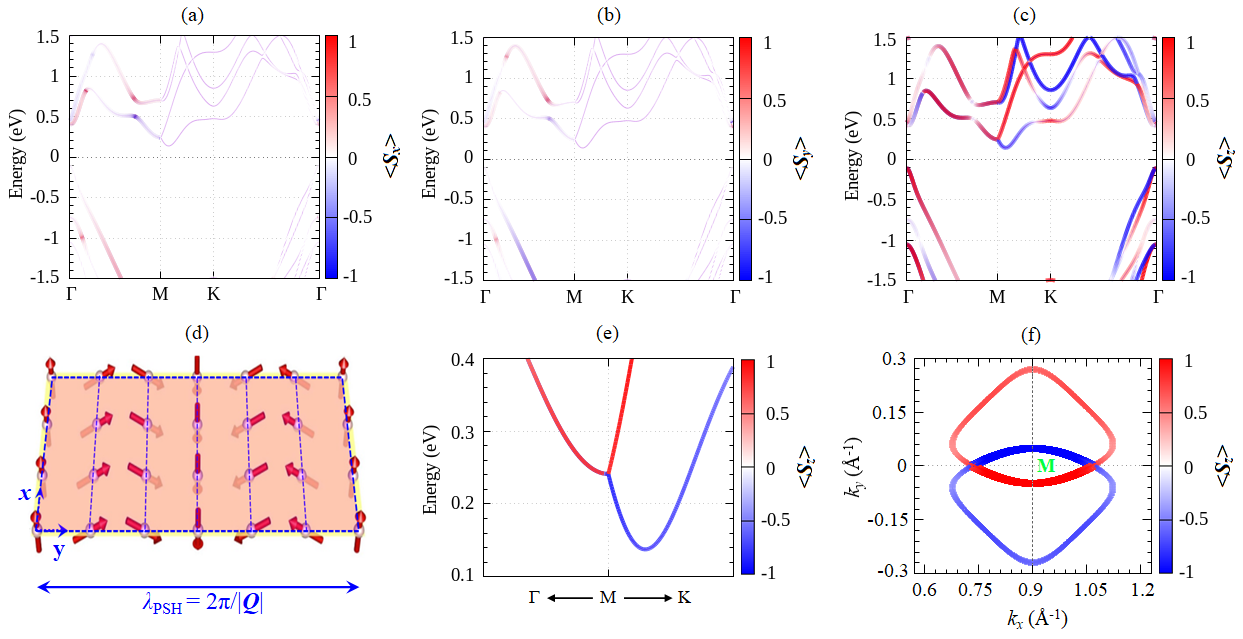}
	\caption{Spin-resolved projected to the bands of the Si$_{2}$Bi$_{2}$ ML for: (a) $S_{x}$,  (b) $S_{y}$, and  (c) $S_{z}$ spin component of the spin polarization $\vec{S}$. (c) Schematic view of the spatially periodic mode of the spin polarization in the persistent spin helix (PSH) state is presented. The wavelength of the PSH, $\lambda_{\texttt{PSH}}$, is indicated by blue arrow. (d) The $S_{z}$ component of spin projected to the bands along the $\Gamma-M-K$ lines calculated around the CBM is presented. (e) The $S_{z}$ component of spin projected to the Fermi line calculated at constant energy cut of 0.1 eV above the degenerated state at the CBM around the $M$ point is shown.}
	\label{fig:Figure3}
\end{figure*}

Figs. 3(a)-(c) shows spin-resolved projected to bands along the $\Gamma-M-K-\Gamma$ line in the FBZ calculated for the Si$_{2}$Bi$_{2}$ ML. We find that all the spin-split bands in the whole FBZ are fully characterized by the out-of-plane spin component ($S_{z}$) [Fig. 3(c)], while the contribution of the in-plane spin components ($S_{x}$, $S_{y}$) is almost zero [Figs. 3(a)-(b)]. This indicates that the spin polarization direction is locked being oriented in the fully out-of-plane direction, leading to the FZPST; consistent with symmetry analysis. We noted here that we have identified the significant spin polarization $\vec{S}(\vec{k})=(S_{x}, S_{y}, S_{z})$ at the degenerated bands along the $\Gamma-M$ line [Figs. 3(a)-(c)]. However, due to the equal population of the spin polarization between the $\vec{S}(\vec{k},\uparrow)$ and $\vec{S}(\vec{k},\downarrow)$ spin-polarized states in the degenerated bands along the the $\Gamma-M$ line, the net spin polarization $\vec{S}$ becomes zero; see Fig. S4 in the Supporting Information \cite{Supporting}. Thus, the FZPST is robust for the non-degenerate bands.   

The emergence of the FZPST in the present system shows that the unidirectional out-of-plane SOF is preserved covering the whole FBZ. In such a situation, electron motion accompanied by the spin precession around the uniaxial SOF lead to a spatially periodic mode of the spin polarization forming a highly stable persistent spin helix (PSH) state \cite{Bernevig, Schliemann, SchliemannJ, Altmann} as schematically shown in Fig. 3(b). Due to $SU(2)$ spin rotational symmetry in the PSH state \cite{Bernevig}, the corresponding spin-wave mode protects the spins of the electrons from dephasing, which is robust against spin-independent scattering, and renders an extremely long spin lifetime by suppressing the DP spin relaxation \cite{Dyakonov}. Therefore, the present systems enable long-range spin transport without dissipation \cite{Altmann}, and hence promising for efficient spintronics devices.

Next, we briefly comment on the main difference between the PST found in the present system compared with that reported in the previous studies. There are two main scenarios to achieve the PST: (i) by fine-tuning the strength of the linear Rashba and Dresselhaus SOCs  \cite{Bernevig, Schliemann, SchliemannJ, Altmann}, or (ii) by imposing the specific CPGS for the systems exhibiting crystallographic polar axis \cite{Tao2018}. Although the former approach was demonstrated on various semiconductor heterostructures  \cite{Koralek2009, Walser2012, Kohda, Sasaki2014}, several problems impede the practical application due to the stringent condition for achieving an equal strength of the Rashba and Dresselhaus SOCs. On the other hand, the latter approach may provide an advantage for realizing the PST since the condition of the equal strength of the Rashba and Dresselhaus SOC can be eliminated. However, in addition to producing the local PST in a small part of the FBZ, this approach is limited only for the systems having polar crystallographic axis as previously reported on several bulk ferroelectric materials including BiInO$_{3}$ \cite{Tao2018}, CsBiNb$_{2}$O$_{7}$ \cite{Autieri2019}, layered ferroelectric oxide Bi$_{2}$WO$_{6}$ \cite{Djani}, and 2D ferroelectric systems such as WO$_{2}$Cl$_{2}$ \cite{Ai2019}, Ga$XY$($X$=Se, Te; $Y$=Cl, Br, I) \cite{Absor2021a, Absor2021b}, hybrid perovskite benzyl ammonium lead-halide \cite {Jia}, and group-IV monochalcogenide \cite{Absor2021c, Absor2019a, Anshory2020, Absor2019b, Lee2020}. In contrast, for our system, there is no crystallographic polar axis observed in the crystal structure, thus the PST is mainly determined by the WPGS rather than by the CPGS \cite{Acosta}. In addition, the PST observed in the present system covers the whole of the FBZ, which is expected to be more beneficial to producing a highly stable PSH state.

\subsubsection{Effective $\vec{k}\cdot\vec{p}$ Hamiltonian.}

To further clarify the emergence of the FZPST, we then focus our attention on the bands around the $M$ point in the CBM due to the large spin splitting as highlighted in Fig. 3(e). Since we have identified the anisotropic nature of the spin splitting around the $M$ point [see Fig. 2(c)], it is expected that the spin-split Fermi line at a constant-energy cut should be highly anisotropic. Our calculated Fermi line measured at a constant-energy cut of 0.1 eV above degenerated states at the $M$ point in the CBM confirmed the anisotropic spin-split Fermi line, where the center of two spin-split Fermi loops are displaced along the $M-K$ ($k_{y}$) direction from their original $M$ point [Fig. 3(f)]. It is also found that these Fermi loops exhibit perfect PST where opposite $S_{z}$ spin components between the inner and outer loops are observed [Fig. 3(f)], which is also consistent with spin-resolved projected to the band along the $\Gamma-M-K$ line shown in Fig. 3(e). These typical Fermi lines as well as the PST observed around the $M$ points are similar to the [110]-Dresselhaus model supporting the PSH state demonstrated previously on the [110]-oriented semiconductor QW \cite{Bernevig} and several 2D systems such as WO$_{2}$Cl$_{2}$ \cite{Ai2019}, Ga$XY$($X$=Se, Te; $Y$=Cl, Br, I) \cite{Absor2021a}, and group IV monochalcogenides \cite{Absor2019b, Lee2020}.  

To analyze the anisotropic spin splitting and PST around the $M$ point, we derive an effective Hamiltonian model using the $\vec{k}\cdot\vec{p}$ invariant method obtained from the symmetry analysis \cite{Winkler, Vajna}. For the particular high symmetry point in the FBZ, the WPGS of the $\vec{k}$ is characterized by a point group $G$, where the matrix representation for the chosen basis functions is given by $\left\{D(\Omega):\Omega\in G\right\}$, where $\Omega$ is the symmetry operations in the WPGS. The derived Hamiltonian should satisfy the following invariant condition:
\begin{equation}
\label{2}
H(\vec{k})=D(\Omega)H(\Omega^{-1}\vec{k})D^{-1}(\Omega), \ \ \forall \Omega\in G.
\end{equation} 

\begin{table}[ht!]
\caption{Transformation rules for the wave vector $\vec{k}$ and spin vector $\vec{\sigma}$ under the $C_{2v}$ point-group symmetry. Time-reversal symmetry, implying a reversal of both spin and momentum, is defined as $T=i\sigma_{y}K$, where $K$ is the complex conjugation, while the point-group operations are defined as $\hat{C}_{2}=i\sigma_{x}$, $\hat{M}_{xz}=i\sigma_{y}$, and $\hat{M}_{xy}=i\sigma_{z}$. The last column shows the invarian terms, where the underlined term are invariant under all symmetry operations.} 
\centering 
\begin{tabular}{c c c c} 
\hline\hline 
  Symmetry  & $(k_{x}, k_{y}, k_{z})$   & $(\sigma_{x}, \sigma_{y}, \sigma_{z})$ & Invariants\\
  Operations  &  &  & \\ 
\hline 
$\hat{T}=i\sigma_{y}K$   &  $(-k_{x}, -k_{y}, -k_{z})$  &  $(-\sigma_{x}, -\sigma_{y}, -\sigma_{z})$  & $k_{i}\sigma_{j}$ ($i,j=x,y,z$)  \\     
$\hat{C}_{2}=i\sigma_{x}$   &  $(k_{x}, -k_{y}, -k_{z})$  &  $(\sigma_{x}, -\sigma_{y}, -\sigma_{z})$ & $k_{x}\sigma_{x}$, $k_{y}\sigma_{y}$, \underline{$k_{y}\sigma_{z}$}, \underline{$k_{z}\sigma_{y}$}, $k_{z}\sigma_{z}$ \\ 
$\hat{M}_{xz}=i\sigma_{y}$   &  $(k_{x}, -k_{y}, k_{z})$  &  $(-\sigma_{x}, \sigma_{y}, -\sigma_{z})$ & $k_{x}\sigma_{y}$, \underline{$k_{y}\sigma_{z}$}, $k_{y}\sigma_{x}$, \underline{$k_{z}\sigma_{y}$}   \\
$\hat{M}_{xy}=i\sigma_{z}$   &  $(k_{x}, k_{y}, -k_{z})$  &  $(-\sigma_{x}, -\sigma_{y}, \sigma_{z})$ &  $k_{x}\sigma_{z}$, \underline{$k_{y}\sigma_{z}$}, $k_{z}\sigma_{x}$, \underline{$k_{z}\sigma_{y}$} \\
\hline\hline 
\end{tabular}
\label{table:Table 3} 
\end{table}

Given that the band dispersion around the $M$ point in the CBM is mainly composed of Bi-$p$ orbitals [Fig. 2(b)], we can ignore other orbitals when constructing the $\vec{k}\cdot\vec{p}$ model. Based on these orbitals, we construct the spinor wave functions to obtain the matrix forms of the symmetry operations in the momentum $\vec{k}$ and spin $\vec{\sigma}$ spaces. Since the WPGS at the $M$ point belongs to $C_{2v}$ point group [see Table 2], both the $\vec{k}$ and $\vec{\sigma}$ can be transformed according to the symmetry operations of this point group. The transformation rules for the $\vec{k}$ and $\vec{\sigma}$ are presented in Table 3. Here, all the invariant terms of the Hamiltonian have listed in the form of product between the $\vec{k}$ and $\vec{\sigma}$ components and then select those specific terms which are invariant under all symmetry operations as indicated by the underlined terms in the right column in Table 3. Finally, we can write the possible term of the Hamiltonian including the Rashba and Dresselhaus terms as
\begin{equation}
H_{M}(k)= H_{0}(k)+\alpha k_{y}\sigma_{z} + \alpha^{'} k_{z}\sigma_{y}
\label{3}
\end{equation}
Here, $H_{0}(k)$ is the Hamiltonian of the free electrons with eigenvalues $E_{0}(k)=\frac{\hbar^{2}k_{x}^{2}}{2m_{x}^{*}}+\frac{\hbar^{2}k_{y}^{2}}{2m_{y}^{*}}$, where $m_{x}^{*}$ and $m_{y}^{*}$ are effective mass of electron evaluated from the band dispersion along the $k_{x}$ and $k_{y}$ directions, respectively. Since our 2D system is lied on the $x-y$ plane, the $k_{z}\sigma_{y}$ term in Eq. (\ref{3}) is naturally disappeared, thus the effective $\vec{k}\cdot \vec{p}$ Hamiltonian can further be expressed as
\begin{equation}
H_{M}=E_{0}(k)+\alpha k_{y}\sigma_{z}
\label{5}
\end{equation}
On diagonalizing Eq. (\ref{5}), we obtain the eigenstates 
\begin{equation}
\Psi_{\vec{k}\uparrow}=e^{i\vec{k}_{\uparrow}\cdot\vec{r}}\left[{\begin{array}{c}
1\\
0
\end{array}}\right]
\label{6}
\end{equation}
and
\begin{equation}
\Psi_{\vec{k}\downarrow}=e^{i\vec{k}_{\downarrow}\cdot\vec{r}}\left[{\begin{array}{c}
1\\
0
\end{array}}\right],
\label{7}
\end{equation}
and corresponding to the eigenenergies
\begin{equation}
E_{M}^{\pm}(k)=E_{0}(k)\pm \alpha k_{y}.
\label{8}
\end{equation}

The spin polarization in the $k$-space can be evaluated from the expectation values of the spin operators, i.e., $\vec{S}=(\hbar/2)\langle \psi_{\vec{k}}|\vec{\sigma}|\psi_{\vec{k}}\rangle$, where $\psi_{\vec{k}}$ is the electron's eigenstates. By using $\psi_{\vec{k}}$  given in Eqs. (\ref{6}) and (\ref{7}), we obtain that \begin{equation}
\vec{S}_{\pm}=\pm \frac{\hbar}{2}[0,0,1].
\label{9}
\end{equation}
This shows that the spin polarization is locked being oriented in the out-of-plane directions, consistent with the spin-resolved projected bands shown in Figs. 3(e).

In addition, we noted here that Eq. (\ref{8}) represents the dispersion of the spin split bands around the original point ($M$ point) exhibiting strongly anisotropic splitting, i.e., the energy bands are lifted along the $k_{y}$ direction ($M-K$ line) but are degenerated along the $k_{x}$ direction ($\Gamma-M$ line), which is consistent with the spin splitting map presented in Fig. 2(c). Interestingly, Eq. (\ref{8}) leads to the following shifting property, $E_{{-}}(\vec{k})=E_{+}(\vec{k}+\vec{Q})$, where $\vec{Q}$ is the shifting wave vector expressed by 
\begin{equation}
\vec{Q}=\frac{2m_{y}^{*}\alpha}{\hbar^{2}}[0,1,0]
\label{10}
\end{equation}
This implies that the band dispersion at a constant-energy cut shows two Fermi loops whose centers are displaced along the $k_{y}$ (M-K line) direction from their original point ($M$ point) by $\pm\vec{Q}$, which is in a good agreement with the Fermi lines shown in Fig. 3(f). Overall, the consistency of the band splitting, Fermi lines, and spin polarization around the $M$ point obtained from the DFT and effective Hamiltonian model justifies the reliability of our method.
 
To further quantify the PST around the $M$ point in the CBM, we here calculate the spin-orbit strength parameter, $\alpha$, by numerically fitting of Eq. (\ref{8}) to the band dispersion along the $\Gamma-M-K$ line obtained from our DFT results. In Table 4, we summarize the calculated values of $\alpha$ for all compounds of the $A_{2}B_{2}$ MLs and compare these results with a few selected PST systems previously reported on several 2D materials. Taking Si$_{2}$Bi$_{2}$ ML as an example, we find that that the calculated $\alpha$ is 2.39 eV\AA, which is comparable with that reported on 2D PST materials including Ga$XY$ ($X$=Se, Te; $Y$=Cl, Br, I) MLs  (0.53 - 2.65 eV\AA) \cite{Absor2021a} and various 2D group IV monochalcogenide such as Ge$XY$ ($X,Y$ = S, Se, Te) MLs (3.10 - 3.93 eV\AA) \cite{Absor2021c}, and layered SnTe (1.28 - 2.85 eV\AA) \cite{Absor2019a, Anshory2020, Absor2019b, Lee2020}. However, this value is much larger than that predicted on WO$_{2}$Cl$_{2}$ ML (0.90 eV\AA) \cite{Ai2019} and transition metal dichalcogenide $MX_{2}$ MLs with line defect such as PtSe$_{2}$ (0.20 - 1.14 eV\AA) \cite{Absor2020} and (Mo,W)$X_{2}$ ($X$=S, Se) (0.14 - 0.26 eV\AA) \cite{Li2019}.

\begin{table*}
\caption{Several selected PST systems in 2D materials and parameters characterizing the strength of the SOC ($\alpha$, in eV\AA) and the wavelength of the PSH mode ($\lambda_{\texttt{PSH}}$, in nm).} 
\begin{tabular}{c c c c} 
\hline\hline 
  2D materials & $\alpha$ (eV\AA)  & $\lambda_{\texttt{PSH}}$ (nm)  & Reference \\ 
\hline 
\underline{$A_{2}B_{2}$ ML compounds} &     &        &    \\
Si$_{2}$Sb$_{2}$ &  0.63   &    11.62    &   This work \\
Ge$_{2}$Sb$_{2}$ &  0.72   &    8.61    &   This work \\
Sn$_{2}$Sb$_{2}$ &  0.80   &    6.50    &   This work \\
Si$_{2}$Bi$_{2}$ &  2.39   &    2.15    &   This work \\
Ge$_{2}$Bi$_{2}$ &  2.41   &    2.13    &   This work \\
Sn$_{2}$Bi$_{2}$ &  2.55   &    2.02    &   This work \\
\underline{Group IV Monochalcogenide} &     &        &    \\
(Sn,Ge)$X$ ($X$= S, Se, Te)  & 0.07 - 1.67 & $8.9\times10^{2}$ - 1.82  & Ref. \cite{Absor2019b}  \\
Ge$XY$ ($X, Y$= S, Se, Te) & 3.10 - 3.93& 6.53 - 8.52  & Ref. \cite{Absor2021c}  \\ 
Layered SnTe & 1.28 - 2.85 & 8.80 - 18.3  & Ref. \cite{Lee2020}  \\ 
Strained SnSe & 0.76 - 1.15 &  & Ref. \cite{Anshory2020}\\
SnSe-$X$ ($X$= Cl, Br, I)  & 1.60 - 1.76 & 1.27 - 1.41 & Ref. \cite{Absor2019a}\\

\underline{Defective 2D TMDCs} &     &        &    \\
line defect in PtSe$_{2}$ & 0.20 - 1.14 & 6.33 -  28.19 & Ref. \cite{Absor2020}\\
line defect in (Mo,W)(S,Se)$_{2}$ & 0.14 - 0.26 & 8.56 - 10.18 & Ref. \cite{Li2019}\\
\underline{Other 2D ML} &     &        &    \\
WO$_{2}$Cl$_{2}$ &  0.90     &      & Ref. \cite{Ai2019}\\
Ga$XY$ ($X$=Se, Te; $Y$=Cl, Br, I) &  0.5 - 2.65     &  1.2 - 6.57    & Ref. \cite{Absor2021a} \\
\hline\hline 
\end{tabular}
\label{table:Table 5} 
\end{table*}

We pointed out here that the observed PST with a large spin-orbit strength parameter, $\alpha$, is expected to induce the highly stable PSH states. Therefore, it is possible to estimate the wavelength of the PSH, $\lambda_{\texttt{PSH}}$, defined as $\lambda_{\texttt{PSH}}=2\pi/\left\|\vec{Q}\right\|$ [see Fig. 3(d)], where the shifting wave vector $\vec{Q}$ is evaluated from Eq. (\ref{10}). The resulting wavelength $\lambda_{\texttt{PSH}}$ for all members of the $A_{2}B_{2}$ ML compounds are listed in Table 4. It is found that the smaller $\lambda_{\texttt{PSH}}$ is achieved for the MLs with larger $\alpha$. In particular, we find a small value of $\lambda_{\texttt{PSH}}$ for Si$_{2}$Bi$_{2}$ ML (2.15 nm), which is smaller than that reported on several 2D group IV monochalcogenide \cite{Absor2021c, Absor2019a, Absor2019b, Lee2020} and defective transition metal dichalcogenides \cite{Absor2020, Li2019} [see Table 4]. Typically, the small wavelength of the PSH state observed in the present system is on the scale of the lithographic dimension achieved by the recent semiconductor industry \cite{Fiori2014}, which is promising for miniaturization and highly scalable spintronics devices.

Thus far, we have found that the FZPST is achieved in the MLs of $A_{2}B_{2}$ compounds. In particular, the PST with the large spin splitting is observed for the Si$_{2}$Bi$_{2}$ ML ($\alpha=2.39$ eV\AA) in the CBM around the $M$ point, indicating
that the PSH state is achieved when the electron carriers are optically injected into the unoccupied state. Moreover, motivated by the observation of the valley Hall effect previously reported on the 2D TMDCs \cite{Mak1489}, it is possible to observe a spin Hall effect in the $A_{2}B_{2}$ ML compounds based on the emergence of the PST. Due to the large spin splitting along the $M-K$ line ($k_{y}$ direction) in the CBM [Figs. 1(c) and 3(c)], the spin polarization is expected to occur in the two states located at the $k_{y}$ and $-k_{y}$ near the CBM; see the spin-resolved projected to the Fermi lines shown in Fig. 3(f). Accordingly, large Berry curvatures with opposite signs are induced. Therefore, it is possible to create an imbalanced population of the spins of electrons in these two states by using polarized optical excitation, and hence a spin Hall current can be detected.

We noted here that to access experimentally the electronic states displaying the PST, the problem of Fermi-level tuning is indispensable due to the semiconductor nature of the electronic state of the $A_{2}B_{2}$ MLs. Here, the application of an external electric field could be a promising method for tuning the Fermi level in the previously reported 2D materials \cite{Ghosh, Ke, Mardanya}. However, introducing the electric field may disturb the stability of the PST due to the breaking of the crystal symmetry \cite{Absor2021c}. In the next subsection, we will show that how the FZPST observed in the present system can be sensitively affected by the external electric field, thus offering a promising platform for spintronic devices such as the SFET.

\subsubsection{The Role of external electric field.}

\begin{figure*}
	\centering
		\includegraphics[width=0.8\textwidth]{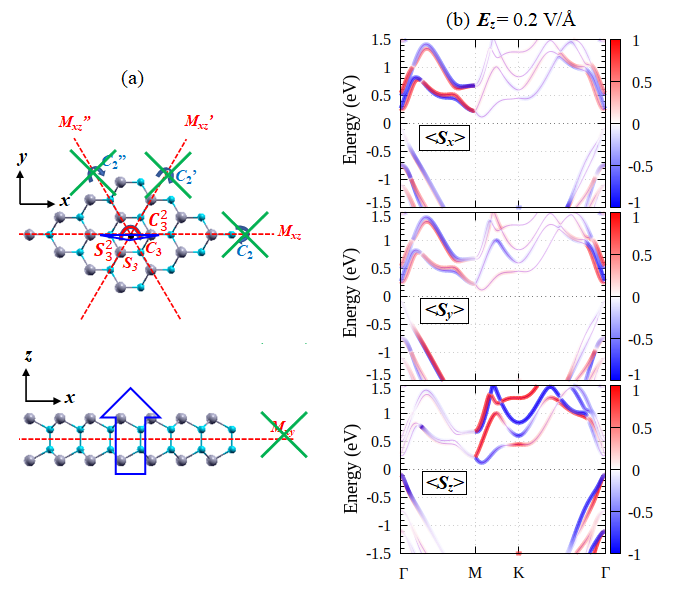}
	\caption{(a) Schematic view of an external out-of-plane electric field applied $E_{z}$ on the $A_{2}B_{2}$ MLs. Here, the external electric field $E_{z}$ is indicated by the blue arrows. The broken of the particular symmetry operations under the influence of the external electric fields are shown by the green cross-lines. (b) Spin-resolved projected to the bands of the Si$_{2}$Bi$_{2}$ ML under $E_{z}$ of 0.2 V/\AA.}
	\label{fig:Figure4}
\end{figure*}

Now, we turn our attention to examining the effect of the external electric field on the FZPST of the $A_{2}B_{2}$ MLs. Here, we introduce the external out-of-plane electric fields, $E_{z}$, oriented along the $\hat{z}$ directions as schematically shown in Fig. 4(a). When the $E_{z}$ is applied, both the $C_{2}$ rotational and $M_{xy}$ in-plane mirror symmetries are broken. The breaking of the particular symmetry operations under the influence of the external electric fields is expected to strongly modify the electronic properties of the $A_{2}B_{2}$ MLs. In fact, an electronic transition from indirect semiconductor to metallic states is observed when the external electric fields are introduced; see Fig. S5 in the Supporting Information \cite{Supporting}, which is attributed to the modulation of the hybridization states between the Bi-$p$, Si-$s$, and Si-$p$ orbitals around the proximity of the band edges (CBM and VBM) near the Fermi level [see Figs. 2(a)-(b)]. The electronic transition states driven by the electric fields observed in the present system are similar to that reported previously on the various 2D materials \cite{Ghosh, Ke, Mardanya}.   

We noted here that the application of the external electric fields also induces modification of the spin-splitting properties of the Si$_{2}$Bi$_{2}$ ML. Fig. 4(b) shows the spin-resolved projected to the bands of the Si$_{2}$Bi$_{2}$ ML under the external electric fields $E_{z}$ of 0.2 V/\AA. Due to the breaking of the $C_{2}$ rotational symmetry of the crystal by the $E_{z}$ [Fig. 4(a)], substantially small band splitting is induced in the bands along the $\Gamma-M$ line, except for the $\vec{k}$ point at the $\Gamma$ and $M$ points due to time reversibility [Fig. 4(b)]. However, the large spin splitting remains for the bands along the $M-K$ line. Since the in-plane mirror symmetry $M_{xy}$ is already broken by the external out-of-plane electric field $E_{z}$, significant in-plane spin components ($S_{x}$, $S_{y}$) are induced in the spin-split bands [Fig. 4(b)], thus breaking the FZPST. Because the large band splitting is located close to the Fermi level, the application of the external electric field may be a hint for future spintronics applications.   

To further analyze the spin-splitting properties of the Si$_{2}$Bi$_{2}$ ML under the out-of-plane external electric fields $E_{z}$, we consider the spin-split polarized bands around the $M$ point in the CBM as highlighted in Fig. 5(a). By adopting the $\vec{k}\cdot\vec{p}$ invariant method \cite{Winkler, Vajna}, the effective Hamiltonian around the $M$ point in the CBM including the electric field contribution can written as     
\begin{eqnarray}
H(k)& =H_{M}(k)+H_{E_{z}}(k)\\
      	& = E_{0}(k)+\alpha k_{y}\sigma_{z} + \alpha_{E_{z}}\left(k_{x}\sigma_{y}+k_{y}\sigma_{x}\right).
\label{11}
\end{eqnarray}
Here, $H_{M}(k)$ is the Hamiltonian without an external electric field, while $H_{E_{z}}(k)$ represents the Hamiltonian induced by an external electric fields, $E_{z}$, characterized by the parameter $\alpha_{E_{z}}=E_{z}\left\langle\Psi(k,r)\left|z\right|\Psi(k,r)\right\rangle$, where $\Psi(k,r)$ is the spinor Bloch wave function of the unoccupied states around the $M$ point in the CBM. By setting the $k_{x}$ and $k_{y}$ axis along the $M-\Gamma$ and $M-K$ lines, respectively, the Hamiltonian of Eq. (\ref{11}) leads to the eigenenergies:
\begin{equation}
E^{\pm}(k)=E_{0}(k)\pm  \sqrt{\left(\alpha_{E_{z}}^{M-\Gamma}\right)^{2}k_{x}^{2}+\left(\alpha_{E_{z}}^{M-K}\right)^{2}k_{y}^{2}}.
\label{12}
\end{equation}
where $\alpha_{E_{z}}^{M-\Gamma}=\alpha_{E_{z}}$, and $\alpha_{E_{z}}^{M-K}=\sqrt{\alpha^{2}+\alpha_{E_{z}}^{2}}$.
We can see that application of $E_{z}$ splits the band degeneracy along the $M-\Gamma$ ($k_{x}$) due to the non-vanishing $k_{x}$-terms in Eq. (\ref{12}), which is consistent with the band dispersion presented in Figs. 4(b). The explicit expressions of spin expectation values for Hamiltonian of Eq. (\ref{11}) can be easily derived from the explicit form of the electron eigenstates, reading
\begin{equation}
\vec{S}_{\pm}= \pm \frac{\hbar}{\delta}\left[\alpha_{E_{z}}^{M-\Gamma}k_{x}, \alpha_{E_{z}}^{M-\Gamma}k_{y}, \alpha k_{y}\right].  
\label{13}
\end{equation}
where $\delta=2\sqrt{\left(\alpha_{E_{z}}^{M-\Gamma}\right)^{2}k_{x}^{2}+\left(\alpha_{E_{z}}^{M-K}\right)^{2}k_{y}^{2}}$. It is clearly seen that introducing $E_{z}$ induces the in-plane spin components ($S_{x}$, $S_{y}$) terms in Eq. (\ref{13}), which is also agree-well with the spin-resolved bands obtained from the DFT results shown in Fig. 4(b).   

\begin{figure*}
	\centering
		\includegraphics[width=0.85\textwidth]{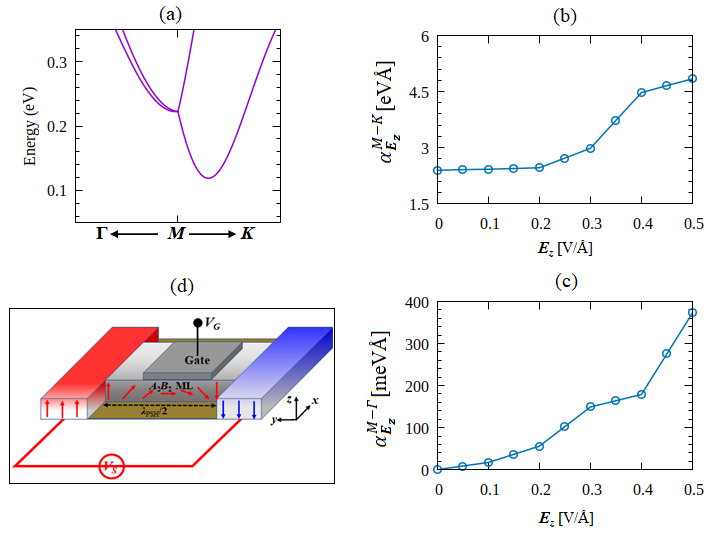}
	\caption{ (a) The spin-split bands around the $M$ point in the CBM under the external out-of-plane electric fields $E_{z}$ of 0.2 V/\AA. Electric field-dependent of the calculated SOC parameters, (b) $\alpha_{E_{z}}^{M-K}$  and (c) $\alpha_{E_{z}}^{M-\Gamma}$ calculated for the spin-split bands along the $M-K$ and $M-\Gamma$ lines around the CBM, respectively. (d) A schematic view of the spin-field effect transistor (SFET) device consisting of ferromagnetic materials in the source and drain parts and the $A_{2}B_{2}$ ML compounds as spin transport channel. The on-off switching logic is controlled by the out-of-plane electric field $\vec{E_{z}}$ through the gate voltage.}
	\label{fig:Figure5}
\end{figure*}

To further characterize the spin-splitting properties of the Si$_{2}$Bi$_{2}$ ML under the influence of the external out-of-plane electric fields, we show in Figs. 5(b)-(c) the electric field-dependent of the SOC parameters ($\alpha_{E_{z}}^{M-\Gamma}$, $\alpha_{E_{z}}^{M-K}$). Here, the $\alpha_{E_{z}}^{M-\Gamma}$ and $\alpha_{E_{z}}^{M-K}$ parameters are obtained by numerically fitting the energy dispersion of the Eq. (\ref{12}) to the DFT bands along the $M-\Gamma$ ($k_{x}$) and $M-K$ ($k_{y}$) lines, respectively, in the CBM around the $M$ point. Interestingly, we observed a strong enhancement of the particular SOC parameters when the external electric fields are introduced. in deed, a sharp increase of more than 100\% occurs for the $\alpha_{E_{z}}^{M-K}$ parameters when the $E_{z}$ are larger than 0.35 V/\AA\ [Fig. 5(b)]. The same trend also holds for the $\alpha_{E_{z}}^{M-\Gamma}$ parameters, see Fig. 5(c). Such a strong enhancement of the SOC parameters by the electric fields helps precisely control the spin precession \cite{WuKai}, which is important for the operation in the SFET device.

Finally, we discuss the possible application of the electric field-driven FZPST in the $A_{2}B_{2}$ ML compounds, which can be implemented as a spin channel in the SFET device. The concept of SFET, the first spintronic device utilizing Rashba SOC, was originally proposed by Datta and Das \cite{Datta}, then various spintronic devices utilizing the PST have been proposed \cite{Schliemann, Absor2021a, Absor2021c, Lee2020}. Here, we propose a similar scheme to design an SFET based on the $A_{2}B_{2}$ ML compounds according to its large SOC parameters and tunability under the electric field, as schematically shown in Fig. 5(d). Similar to the methods employing semiconductor QW\cite{Kohda_a}, the on-off logical functionality of the SFET can be performed by a purely electric manipulation of the anisotropy spin lifetime due to the presence or absence of the PSH state. In the absence of the out-of-plane electric field ($E_{z}=0$), the out-of-plane orientation of the spin polarization injected from the ferromagnetic (FM) source electrode maintains due to the robust FZPST. Here, the spin polarizations are always parallel or anti-parallel to those in the FM drain electrode depending on the electron’s momentum in the specific sub-bands. In contrast, applying $E_{z}\neq 0$ subsequently breaks the FZPST and hence perturbs the PSH state. Accordingly, the spin dephasing may occur due to the spin scattering, thus significantly reducing the spin currents detected by the FM drain electrode. Since the wavelength $\lambda_{\texttt{PSH}}$ of the PSH state observed in the present systems is very small, for an instant, $\lambda_{\texttt{PSH}}=2.15$ nm for the Si$_{2}$Bi$_{2}$ ML, the very small spin channel of the SFET can be realized, which is in favor of the preservation of spin coherence and can be integrated into nanodevices with higher density.

\section{Conclusion}

In summary, based on first-principles DFT calculations supported by symmetry analysis, we systematically investigated the SOC-related properties of the group IV-V $A_{2}B_{2}$ ($A$ = Si, Sn, Ge; $B$ = Bi, Sb) ML compounds. We found a phenomenon called the FZPST, i.e., a unidirectional spin polarization occurring in the whole FBZ. We found that this FZPST observed in the spin-split bands exhibits the fully out-of-plane spin polarization in the whole FBZ, which is enforced by the in-plane mirror symmetry $M_{xy}$ operation in the WPGS for the $\vec{k}$ in the whole FBZ. Importantly, we identified giant spin splitting in which the FZPST sustains, which is particularly visible in the proximity of the CBM, thus supporting the large SOC parameters and small wavelength of the PSH states. Furthermore, our $\vec{k}\cdot\vec{p}$ analysis based on the symmetry argument demonstrated that the FZPST is robust for the non-degenerate bands, which can be effectively controlled by the external out-of-plane electric field. Therefore, we proposed a design of SFET based on the $A_{2}B_{2}$ ML compounds, which have a short spin channel length and can electrically control spin precession efficiently.

Since the FZPST found in the present study is solely enforced by the in-plane mirror symmetry $M_{xy}$ operation in the WPGS for the arbitrary $\vec{k}$ in the whole FBZ, it is expected that this PST can also be achieved on other 2D materials having the similar WPGS. Our symmetry analysis clarified that the following CPGS of the 2D systems are found to support the similar WPGS maintaining the FZPST, including $C_{s}$, $C_{2v}$, $C_{3h}$, and $D_{3h}$. Therefore, our prediction is expected to trigger further theoretical and experimental studies to find novel 2D systems supporting the FZPST, which is useful for future spintronic applications.

\ack{This work was supported by PD Research Grants (No.1709/UN1/DITLIT/Dit-Lit/PT.01.03/2022) funded by KEMDIKBUD-DIKTI, Republic of Indonesia. The computation in this research was partly performed using the computer facilities at Universitas Gadjah Mada, Republic of Indonesia.}

\section*{References}
\bibliography{Reference}

\end{document}